\documentclass[12pt]{article}
\usepackage{amsmath,amsfonts,amssymb}
\usepackage{graphicx} 
\usepackage{bbm}
\usepackage{bm}

\usepackage[utf8]{inputenc}
\usepackage[hypertexnames=false]{hyperref}

\title{Comment on \lq{}Observation of a quantum Cheshire Cat in a matter-wave interferometer experiment\rq{}, Nature Comm. \textbf{5}, 4492}
\author{Antonio Di Lorenzo\\
Instituto de F\'{\i}sica, Universidade Federal de Uberl\^{a}ndia,\\ 38400-902 Uberl\^{a}ndia, Minas Gerais, Brazil}
\begin{document}
\maketitle
\begin{abstract}
It is shown that a classical experiment using an ordinary cat can reproduce the same results and it is argued that the quantum nature of the phenomenon could be revealed instead by making an experiment that detects cross-moments. 
\end{abstract}
\maketitle
In the recent paper \cite{P1}, the authors claim to have separated the mass of a neutron from its polarization, implementing the proposal of Aharonov \emph{et al.}~\cite{P}.
We prove that the experimental data do not support the claim, as they could be reproduced classically, but that, with a suitable change of the apparatus, it would be possible to 
prove uncontroversially that a neutron inside an interferometer is simultaneously in the two arms.

Let us make a classical experiment. 
I send my cat to a bifurcation. I measure her passage in either direction with two very noisy apparatuses. 
In each repetition, either apparatus 
can give me outputs like -200, or +10, or 3.14 cats have passed here, because of a large random bias in the initial position of the pointer. Precisely, if the cat is present, the pointer is shifted by 1, but the initial value $x_0$ of the pointer is not zero, as one would expect 
from an ordinary measurement; rather, it is distributed around 0 with a widespread probability distibution. 
I repeat the measurements $N$ times. 
Then, I change the apparatus, so that now it measures how many times a tail pointing up or down is present in either path. Again, the apparatuses are very noisy. 
I repeat another $N$ times. 
The first average gives me that there is 1 cat in the left path and 0 cats on the right path. The second average yields 0 tails in the left path and 1 tail in the right path. 

Can I claim that I have separated the tail (or, better, the direction of the tail) from my cat (without hurting the cat)? 
Given that the random bias in the apparatuses averages to zero, I would conclude so,  if the number of  repetitions is sufficiently large and \emph{if I used all of the experimental data, i.e. if I did not make a post-selection.} 
However, the post-selection can yield a biased subset of data. 

Indeed, let us say that after the measurement my cat can either eat her food or hide inside a box. The post-selection is considered successful when the cat hides inside the box. 
The initial preparation of the cat consisted in letting her starve for some time. Therefore, most of the times the cat will choose the food over the box, with the exception of a few cases when the cat demonstrates her volatile nature. 
Now, with a small probability, the measuring apparatus can make a scary noise that will cause the cat to hide inside the box, the fear prevailing over the hunger. 
This probability is a function of the initial random bias of the probe. As the probability of the scary noise being produced is always small, the measurement apparatus 
does not disturb significantly the state of the cat. However, the rules of probability theory establish that the subset of events in which the post-selection was successful 
correspond to an initial bias of the detector close to the maximum of the likelihood $prob(noise|amount\ of\ bias)$. 
Therefore, if I consider only the subset of events with a successful post-selection, I can have an average number of cat in the left path equal to 1, and the same average number of 
tail-up in the right path. This does not allow me to conclude that my experimental data showed a disembodiment of the tail from the cat. 
This \emph{gedankenexperiment} can be performed with a classical system, and reproduce the situation that the authors of the article have realized with neutrons.  

Differently from the cat in the above example, the neutron can be thought of as simultaneously present in both arms of the interferometer. However, measuring weakly its average presence and its average polarization, as was done in the paper, provides no evidence of this, as we just demonstrated. 
Instead, we have suggested elsewhere \cite{P2,P3} that a good quantifier of this ubiquity is the sum of the signed cross-moment $\sum_{\pm}\pm\langle xy\rangle_{\pm}$, 
where $x$ is the output of the detector measuring the presence of the neutron in the left arm and $y$ it the output of the detector measuring \emph{simultaneously} (and not in a separate run, as was done in the experiment) the polarization in the right arm, while the sign $\pm$ denotes a successful or unsuccessful post-selection. 
This quantity was shown to be related to the entanglement between the two detectors, which is created due to the neutron being simultaneously in both arms. 

In conclusion, the experiment does not demonstrate a quantum Cheshire cat yet, but it could do so with some appropriate changes. 
%
\section*{Quantifying our thesis}
We provide an explicit example. 
The cat enters either path of the bifurcation with probability $1/2$, and its tail is either up or down with probability $1/2$, irrespectively of the path chosen. 
The cat detector has an initial distribution of the pointer variable 
\begin{equation}
\Pi_0(x) = \frac{1}{\sqrt{2\pi}\Delta} \exp[-x^2/2\Delta^2],
\end{equation}
with $\Delta\gg 1$. The pointer $x$, when the cat chooses the left path, is shifted deterministically from its initial value by $1$, otherwise it stays the same.  
Analogously, in the right path we put a tail detector, with an initial distribution
\begin{equation}
\Pi_0(y) = \frac{1}{\sqrt{2\pi}\Delta} \exp[-y^2/2\Delta^2]. 
\end{equation}
 If the cat chooses the right path and its tail is up, $y$ shifts deterministically by $+1$, it its tail is down, $y$ shifts by $-1$, and if the cat is not there, $y$ does not vary. 
After the measurement, if the value of the pointer is $x$, the cat detector makes a scary noise with the probability 
\begin{equation}
p_c(x)\equiv P_c(noise|x) = \varepsilon_c \exp[-(x-u)^2/2\delta^2],
\end{equation}
with $\varepsilon_c\ll 1$, and $u$, $\delta$ parameters that we shall fix appropriately. 
Analogously, the tail detector can produce a noise with probability 
\begin{equation}
p_t(y)\equiv P_t(noise|y) = \varepsilon_t \exp[-(y-v)^2/2\delta^2].
\end{equation}
The detectors are acustically insulated from each other, so that if the cat detector makes a noise but the cat is interacting with the tail detector, the cat won't hear any noise. We are assuming this to exclude a non-local feedback process. 
After the measurement, if no noise was produced, the cat will either eat its food or hide in the box, with a probability $q=3/4$ and $p=1/4$, respectively. 
However, in the rare cases when a noise is produced in the proximity of the cat, she will hide in the box with probability $1$. 

We can calculate the joint probability of observing an output $x$ from the detector and later post-selecting the cat inside the box (which we denote by the symbol $b$) 
by applying Bayes' rule
\begin{align}
\nonumber
\Pi(x,y,b) =& 
\int dx_0 dy_0 \Pi(x,y,b|x_0,y_0) \Pi_0(x_0,y_0) \\
\nonumber
=&\int dx_0 dy_0 P(b|x,y,x_0,y_0) \Pi(x,y|x_0,y_0) \Pi_0(x_0,y_0) \\
=& 
\int dx_0 dy_0 P(b|x,y,x_0,y_0)\Pi(x,y|x_0,y_0) \Pi_0(x_0)\Pi_0(y_0) 
,
\label{main}
\end{align}
where, in the last line, we used the independence of 
the initial values of $x$ and $y$, $\Pi_0(x_0,y_0)=\Pi_0(x_0)\Pi_0(y_0)$. 
Now, the probability density of observing a final shift $x$, given that the initial value of the pointer was $x_0$, is obtained by considering the two classical alternatives: 
either the cat went left, thus increasing $x_0$ by 1 and leaving $y_0$ to its original value, or the cat went right, leaving $x_0$ untouched and either increasing or decreasing $y_0$ by one unit. 
Namely, 
\begin{equation}
\Pi(x,y|x_0,y_0) =\frac{1}{2}\delta(x-x_0-1)\delta(y-y_0)
 +\frac{1}{4}\delta(x-x_0)\delta(y-y_0-1)+\frac{1}{4}\delta(x-x_0)\delta(y-y_0+1).
\end{equation}
On the other hand, the conditional probability of post-selection is 
\begin{align}
\nonumber
P(b|x,y,x_0,y_0)=&\sum_{n_c,n_t=0}^1 P(b,n_c,n_t|x,y,x_0,y_0)\\
\nonumber
=& \sum_{n_c,n_t=0}^1 P(b|n_c,n_t,x,y,x_0,y_0) P(n_c,n_t|x,y,x_0,y_0)
\\
=& P(b|1) p_c(x)p_t(y)+\left[P(b|1)\delta_{y,y_0}+P(b|0)\delta_{x,x_0}\right] p_c(x)[1-p_t(y)]
\nonumber\\
&
+\left[P(b|0)\delta_{y,y_0}+P(b|1)\delta_{x,x_0}\right] [1-p_c(x)]p_t(y)+P(b|0) [1-p_c(x)[1-p_t(y)] 
\end{align}
where $n_c,n_t=0$ indicates that no scary noise was produced, and $n_c,n_t=1$ that the detector made the scary noise. 
In the last equality, we exploited the fact that the probability of the cat going into the box depends only on her hearing the noise, 
all the other parameters being superfluous, and 
we considered that it may happen that a detector makes a noise when the cat is not there, hence the Kronecker deltas. 
We recall that we defined $p=P(b|0)$ the probability that the cat will go inside the box when undisturbed; 
on the other hand $P(b|n=1)=1$. 

Finally, substituting into Eq.~\eqref{main}, we have 
\begin{align}
\nonumber
\Pi(x,y,b) &=\frac{1}{2} \Pi_0(x-1)\Pi_0(y)
\{p[1-p_c(x)]+p_c(x)\}
\nonumber
\\
&
+\frac{1}{4}\Pi_0(x)\left[\Pi_0(y-1)+\Pi_0(y+1)
\right]
\{p[1-p_t(y)]+p_t(y)\}.
\label{joint}
\end{align}
The conditional average outputs are then 
\begin{equation}
\langle x\rangle_b = \frac{\int dx dy\, x \Pi(x,y,b)}{{\int dx dy \Pi(x,y,b)}} 
\end{equation}
and 
\begin{equation}
\langle y\rangle_b = \frac{\int dx dy\ y \Pi(x,y,b)}{{\int dx dy \Pi(x,y,b)}}.
\end{equation}
The denominator in these expressions is the probability of post-selection
\begin{align}
P(b)&=\int dx dy \Pi(x,y,b) 
\nonumber
\\
&= p+\frac{q}{4} \frac{\Delta_R}{\Delta} 
\left\{2\varepsilon_c G(u-1)+\varepsilon_t\left[ G(v-1)+G(v+1)\right]\right\}
\nonumber
\\
\nonumber
&\simeq p+O(\varepsilon) =\frac{1}{4}+O(\varepsilon),
\end{align}
where we introduced the reduced variance 
\begin{equation}
\Delta^2_R = \frac{\Delta^2\delta^2}{\Delta^2+\delta^2}, 
\end{equation}
and we defined 
\begin{equation}
G(x) = e^{-x^2/[2(\Delta^2+\delta^2)]}.
\end{equation}

The post-selected averages, after some straightforward albeit tedious calculations, are 
\begin{align}
\langle x\rangle_b \simeq& \frac{1}{2} + \frac{3}{2} \frac{\Delta_R}{\Delta} 
\varepsilon_c u'(1)G(u-1)
\label{mainav1}
\end{align}
and 
\begin{align}
\langle y\rangle_b \simeq& \frac{3}{4} \frac{\Delta_R}{\Delta} 
\varepsilon_ t\left[v'(1)G(v-1)+v'(-1) G(v+1)\right]
,
\label{mainav2}
\end{align}
where we introduced the notation
\begin{equation}
u'(z) = \frac{ \delta^2 z+\Delta^2 u}{\Delta^2+\delta^2}
\end{equation}
and
\begin{equation}
v'(z) = \frac{ \delta^2 z+\Delta^2 v}{\Delta^2+\delta^2} . 
\end{equation}

Inspection of Eqs.~\eqref{mainav1} and \eqref{mainav2} reveals that the correction to the average values, 
even though they are proportional to $\varepsilon_c\ll 1,\varepsilon_t\ll 1$, can be non-negligible if $u,v\sim \Delta$ and if  
$1/\varepsilon_{j}\lesssim\delta\ll \Delta$. 
For instance, we are able to reproduce the theoretical values $\langle x\rangle_b=1$, $\langle y\rangle_b=1$, $P(b)\simeq 0.251$, choosing 
$\Delta=1000$, $\delta=10^{-2}\Delta$, $\varepsilon_c=e/3\delta\simeq 0.091$, $\varepsilon_t=2e/3\delta\simeq 0.181$, $u\simeq 0.402\Delta$, $v\simeq 0.400\Delta$. 
On the other hand, the cross-moment $\langle xy\rangle_b$ for this classical model is 0, while the corresponding quantum mechanical 
realization has, according to the theory, a non-zero value. 
This concludes the proof that a classical experiment with post-selection can reproduce the quantum Cheshire cat, as far as local averages are concerned.

%
\section*{Acknowledgments}
This work was performed as part of the Brazilian Instituto Nacional de Ci\^{e}ncia e
Tecnologia para a Informa\c{c}\~{a}o Qu\^{a}ntica (INCT--IQ).

\end{document}